\documentclass[final,5p,times,twocolumn]{elsarticle}

\usepackage{lineno}
\usepackage[hidelinks]{hyperref}
\usepackage{graphicx}
\usepackage{amsmath}   
\def\pmbanner{{\hrule height 1 pt}\vskip35pt{NIMA POST-PROCESS BANNER TO BE REMOVED AFTER FINAL ACCEPTANCE}\vskip35pt{\hrule height 4pt}\vskip20pt}
\begin{document}

\begin{frontmatter}

%% Note: \pmbanner before the actual title
\title{\pmbanner Performance of small-diameter muon drift tube chambers with new fast readout ASIC at high background rates }

\author[1]{Sergey Abovyan}

\author[1]{Nayana Bangaru}

\author[1]{Francesco Fallavollita}

\author[1]{Oliver Kortner}

\author[1]{Sandra Kortner}

\author[1]{Hubert Kroha}

\author[1]{Elena Voevodina\corref{cor1}%
  }
\ead{voevodin@mpp.mpg.de}

\author[1]{Robert Richter}

\author[1]{Yazhou Zhao}

 \cortext[cor1]{Corresponding author}
% \fntext[fn1]{This is the first author footnote.}
\affiliation[1]{organization={Max-Planck-Institute for Physics}, 
                 addressline={Boltzmannstr. 8},
                 postcode={85748}, 
                 city={Garching}, 
                 country={Germany}}

\begin{abstract}
 Experiments like ATLAS at the HL-LHC or detectors at future hadron colliders need muon detectors with excellent momentum resolution at the percent level up to the TeV scale both at the trigger and the offline reconstruction level. This requires muon tracking chambers with high spatial resolution even at the highest background fluxes. Drift-tube chambers are the most cost effective technology for the instrumentation of large-area muon systems providing the required high rate capability and three-dimensional spatial resolution. Thanks to the advances in analog and digital electronics, the new generation small-diameter Muon Drift Tube (sMDT) detectors with 15 mm tube diameter can be used in stand-alone mode up to the background rates as high as expected at future hadron collider experiments, providing event times and second coordinates without the necessity of additional trigger chambers. New key developments in the integrated front-end electronics are fast baseline restoration of the shaped signal and picosecond time-to-digital converters for second coordinate measurement with double-sided read-out of the tubes. Self-triggered operation has become possible using modern high-performance FPGAs allowing for real-time pattern recognition and track reconstruction. A new amplifier shaper discriminator chip in 65 nm TSMC CMOS technology with increased sensitivity and faster baseline recovery has been developed to cope with very high background fluxes. Extensive test beam campaign using sMDT chamber equipped with new readout electronics has been performed at the CERN Gamma Irradiation Facility (GIF++). The results which will be discussed in this contribution shown that thanks to the shorter peaking time of the new chip, in comparison to its predecessor, leads to an enhancement in the spatial resolution of the drift tubes by up 100 $\mu$m to up to a background rate of 1 MHz which is the maximum rate expected at the 100 TeV collider experiment.

%Experiments like ATLAS at the HL-LHC and future hadron colliders require muon detectors with excellent momentum resolution up to the TeV scale, necessitating high spatial resolution even at high background fluxes. Drift-tube chambers are cost-effective for large-area muon systems, providing high rate capability and three-dimensional spatial resolution. The new generation small-diameter Muon Drift Tube (sMDT) detectors, with 15 mm tubes, can operate in stand-alone mode at high background rates, eliminating the need for additional trigger chambers. Advances in integrated front-end electronics, including fast baseline restoration and picosecond time-to-digital converters, enhance performance. Modern high-performance FPGAs enable self-triggered operation, real-time pattern recognition, and track reconstruction. A new 65 nm TSMC CMOS amplifier shaper discriminator chip, with increased sensitivity and faster baseline recovery, has been developed to handle high background fluxes. Test beam campaigns at the CERN Gamma Irradiation Facility (GIF++) demonstrate that the new chip improves spatial resolution up to a background rate of 1 MHz, the maximum expected at future 100 TeV collider experiments.

\end{abstract}

\end{frontmatter}

%\maketitle
%\pagenumbering{gobble}

% Note that keywords are not normally used for peer review papers.
%\begin{IEEEkeywords}
%IEEE, IEEEtran, journal, \LaTeX, paper, template.
%\end{IEEEkeywords}

\section{High background rates at Future Circular Hadron Collider (FCC-hh)}
In the conceptual design of a detector for a future 100 TeV pp collider (FCC-hh) \cite{b1}, the muon system surrounds the inner detector's solenoid magnet and consists of four parts:
  \begin{itemize}
    \item $|\eta|$ $<$ 1.0: Barrel muon system
    \item 1.0 $\leq$ $|\eta|$ $<$ 1.5: Outer end-cap muon system
    \item 1.5 $\leq$ $|\eta|$ $<$ 2.2: Inner end-cap muon system
    \item 2.2 $\leq$ $|\eta|$ $<$ 3.0: Forward muon system
  \end{itemize}
Different background hit fluxes are expected in each $\eta$-region, with counting rates up to 500 Hz/cm$^2$ in the barrel and outer end-cap regions, and up to 10 kHz/cm$^2$ or 25 kHz/cm$^2$  with a safety factor of 2.5 in the inner end-cap and forward regions.

The muon detector system has two essential functions: providing a muon trigger and achieving $\sim$ 10\% momentum resolution at 10 TeV muon energy by precise muon track reconstruction with an angular resolution of 70 $\mu$rad. This precision is possible with small-diameter drift-tube chambers (sMDT), which can operate effectively at rates up to 30 kHz/cm$^2$ with 30\% occupancy. sMDT technology, featuring two multilayers of four 15 mm diameter drift tube layers spaced 1.5 m apart, can ensure an angular resolution better than 70 $\mu$rad and with the single tube resolution below 150 $\mu$m. \cite{b2}.

\section{sMDT detector design and high rate effects}
\label{sec2}

Aluminum sMDT drift tubes have a diameter of 15 mm and a wall thickness of 0.4 mm. They are filled with an Ar/CO$_2$ (93/7) gas mixture at 3 bar absolute and are operated at 2730 V, resulting in a nominal gas amplification factor G of 2 x 10$^4$ \cite{b3}.

The degradation of spatial resolution and detection efficiency in sMDT detectors at high background hit rates is caused by the following effects \cite{b3}:
  \begin{itemize}
 \item \textit{Increased Occupancy}: Making it difficult to distinguish individual muon hits from background hits and reducing the spatial resolution due to the pile-up of $\gamma$ background and muon hits.
 \item \textit{Dead Time}: Higher likelihood of missed muon events due to new hits occurring during the detector's dead time, decreasing detection efficiency.
 \item \textit{Gas Gain Drop and Fluctuations}: The presence of positive space charge within a tube from ions created by background hits leads to a reduction of the electric field at the anode wire, hence to a gain drop. This drop can be compensated by increasing the operating voltage. Gain fluctuation caused by fluctuations of the space charge has a negligible impact on the tube's performance.
  \end{itemize}
In order to mitigate these effects, the sMDT detector must be equipped with improved front-end electronics featuring fast baseline restoration of the shaped signal to handle high rates and reduce dead time.

\section{New 65 nm ASD chip}
\label{sec3}

A four-channel Amplifier/Shaper/Discriminator (ASD) chip, developed by the Max Planck Institute for Physics (Munich) and fabricated using 65 nm TSMC CMOS technology, has been designed to enhance performance at high counting rates in sMDTs. 

\begin{figure}[htbp]
\centerline{\includegraphics[width=7.0cm]{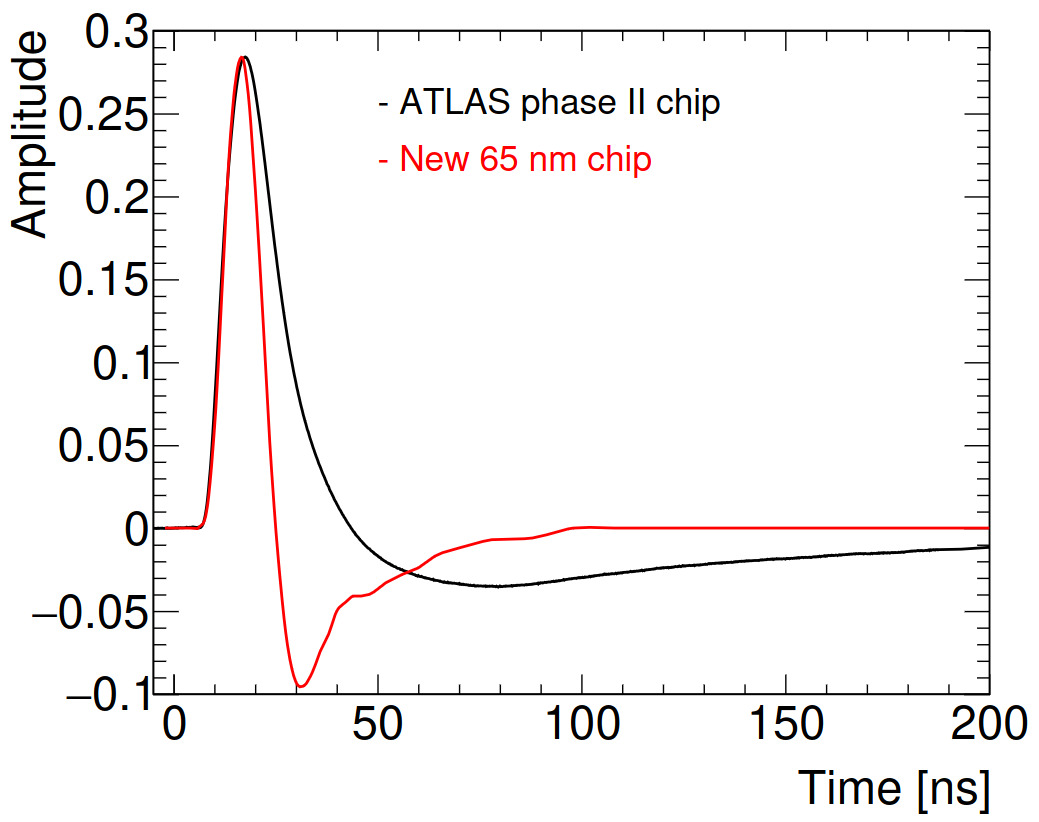}}
\caption{Comparison of the $\delta$ response functions of the new ASD chip and the old ASD chip currently used on ATLAS MDT chambers.}
\label{fig1}
\end{figure}

The ASD chip features bipolar shaping to reduce the effects of baseline shift at high signal rates and has a much faster baseline recovery than the ATLAS ASD chip (Figure \ref{fig1}), reducing the signal pile-up effect. It provides output as both low voltage differential signals and digital CMOS level signals. Each channel consumes 12.8 mW of power, which is 61.2\% less than the ATLAS ASD chip used in the HL-LHC phase II. Additionally, each channel occupies just 0.235 mm$^2$, which is 43\% of the area of the current ATLAS ASD chip.

\section{Performance of the sMDT detector instrumented with the new 65 nm ASD chip at high background rates}
\label{sec4}

The performance of the new chip was tested on an sMDT chamber at the CERN Gamma Irradiation Facility (GIF++) under gamma background irradiation with a high-energy muon beam and compared with the ATLAS chip.

\begin{figure}[htbp]
\centerline{\includegraphics[width=7.0cm]{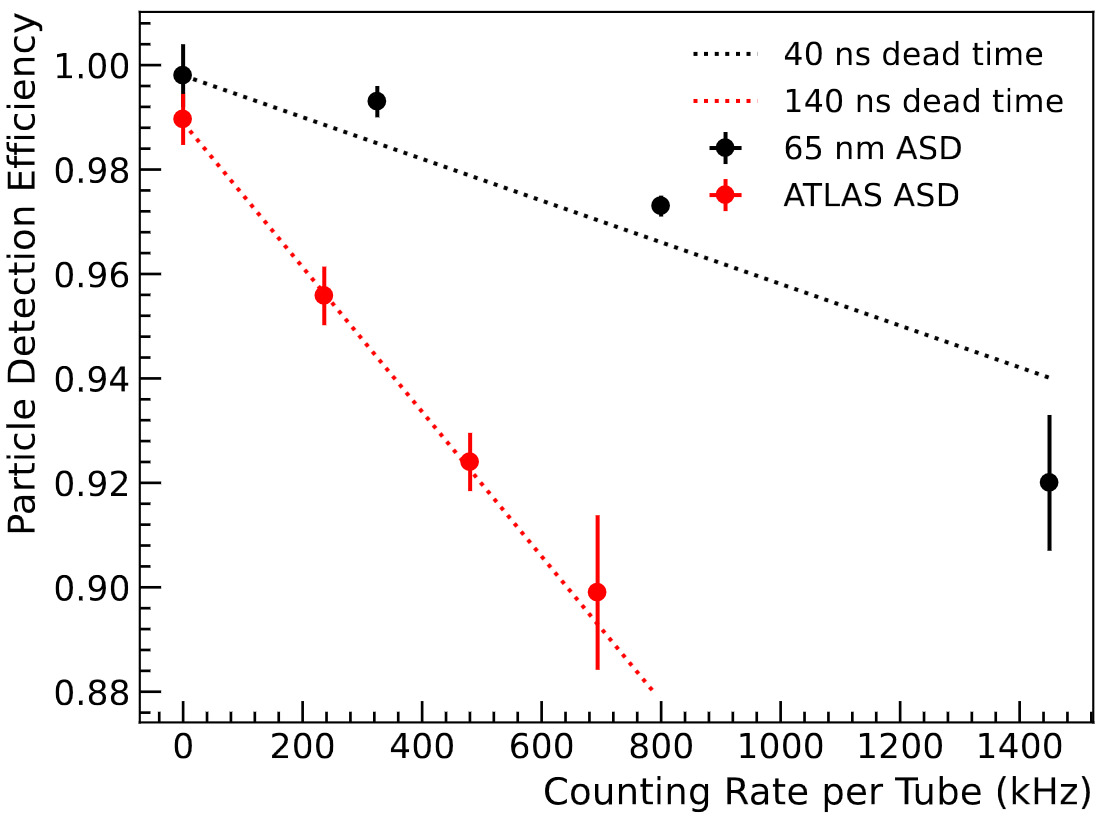}}
\caption{Particle detection efficiency as a function of the average counting rate per tube in kHz without correction for multiplicity. Line of 40 ns dead time (black dashed line) and 140 ns (red dashed line) shown. The dead time of the 65 nm ASD is nearly four times lower than the ATLAS ASD.}
\label{fig2}
\end{figure}

Figure \ref{fig2} shows the study of the muon detection efficiency for 65 nm ASD and ATLAS ASD chips at the different background counting rate. Without $\gamma$ background the muon detection efficiency is less than 100\%, namely (99.0$\pm$0.6)\% . The efficiency decreases with increasing $\gamma$ count-rate due to the increasing probability that a $\gamma$ hit masks a muon hit within the dead time of the front-end electronics \cite{b5}. The efficiency dependence on the $\gamma$ background hit rate is compatible with the expectation for 140 ns dead time of the ATLAS ASD chip and 40 ns dead time of the new 65 nm ASD chip, which was achieved by its fast baseline recovery. As a consequence of the small dead time, a very high muon detection efficiency of $>$ 92\% at 1.4 MHz background hit rate is observed.

\begin{figure}[htbp]
\centerline{\includegraphics[width=7.0cm]{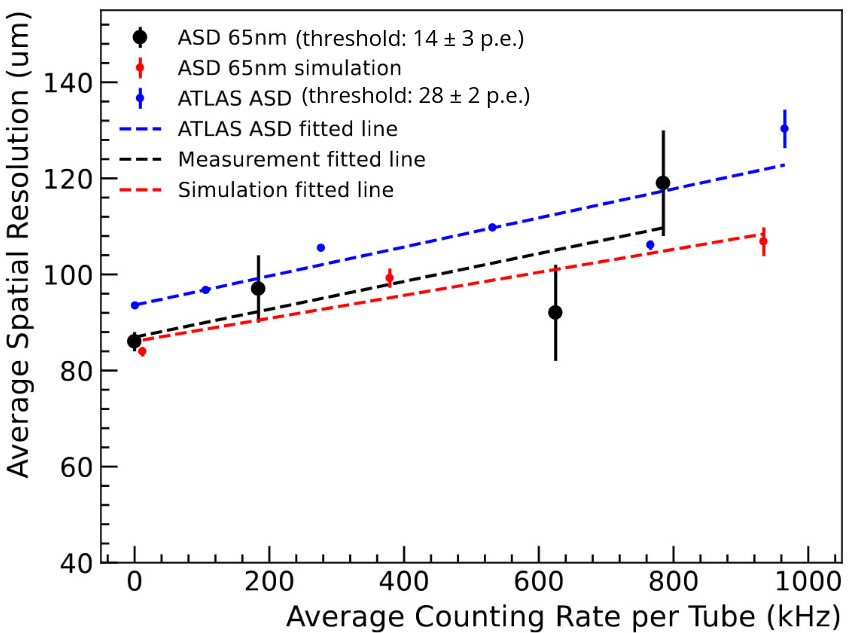}}
\caption{Spatial resolution of the 65 nm ASD from the test beam measurements (black), the Garfield++ simulation (red) and the ATLAS ASD from the test beam measurements (blue) under irradiation. The dashed lines show the fit to the resolution points.}
\label{fig3}
\end{figure}

Figure \ref{fig3} summarizes the dependence of the average spatial resolution of an sMDT on the $\gamma$ count-rate per tube. The sMDT spatial resolution with the 65 nm ASD chip is compatible with the simulation and is better by 10 $\mu$m than with the current ATLAS ASD, 85 $\mu$m without irradiation, and it is expected to be around 110 $\mu$m for a counting rate of 1 MHz/tube. The degradation of the spatial resolution is totally caused by the pile-up of muon and background hits \cite{b5}.

\section{Conclusion}
The new 65 nm ASD chip significantly enhances the performance of sMDT chambers under high background rates, providing improved spatial resolution and detection efficiency, making it suitable for future high-energy physics experiments.

\end{document}